# High-performance graphene-based electrostatic field sensor

Wenhui Wang[1], Ruxia Du[2], Linping He[1], Weiwei Zhao[3], Yunfei Chen[3], Junpeng Lu[1] and Zhenhua Ni[1,*]

[1]Department of Physics and Key Laboratory of MEMS of the Ministry of Education, Southeast University, Nanjing 211189, China.

[2]Department of Basic Teaching, Nanjing Tech University Pujiang Institute, Nanjing 211134, China.

[3]Jiangsu Key Laboratory for Design and Fabrication of Micro-Nano Biomedical Instruments, School of Mechanical Engineering, Southeast University, Nanjing 211189, China.

**Abstract:** Electrostatic sensing technology is widely utilized in both military and civilian applications, including electrostatic prevention in gas stations and various electronic devices. The high sensitivity of electrostatic sensor is capable to detect not only weak electrostatic charges, but also the weak disturbance of electrostatic field in distant. Here, we present a high-performance graphene-based electrostatic sensor. Combining the ultrahigh mobility of graphene and the long lifetime of carriers in lightly doped $SiO_2$/Si substrate, our device achieves a fast response of ~2 μs and detection limit of electrostatic potential as low as ~5 V, which is improved by an order of magnitude as compared to commercial product. The proposed device structure opens a promising pathway to high-sensitive electrostatic detection, and also greatly facilitates the development of novel sensors, e.g. portable and flexible electrostatic sensor.

**Correspondence and requests for materials should be addressed to Z. N. (zhni@seu.edu.cn).**

Static electricity exists widely in moving charged objects due to a variety of electrification process[1-3]. Electrostatic sensing technology has received much attention and is widely applied in both military and civilian fields[1-9], e.g., electrostatic prevention in gas stations or electronic devices. In particular, high-speed flying aircraft inevitably carry a large amount of charges, which cause an electrostatic field that can be detected in kilometers away. Remote detection requires excellent sensitivity of the electrostatic sensor. In addition, the sensor needs to be miniaturized or integrated under certain circumstances, such as portable devices and microelectronic devices. Due to the large leakage resistance and the weak charge signal, the detection limit as well as sensitivity of current sensor is not high, and it sometimes needs amplifier circuit, which would increase the complexity and also the cost of the manufacture process. Therefore, the development of high performance electrostatic sensors with new concept and materials are highly desired.

The emergence of novel two-dimensional materials[10-12] provides an opportunity for ultrasensitive sensors with excellent performance. For instance, graphene is an ideal candidate for high performance sensors or detectors due to its high mobility[13]. Sensitive gas detectors[14,15], molecular probe[16], biological sensor[17,18], and PH sensor[19] based on graphene have been demonstrated. Here, we present a high-performance graphene based electrostatic sensor on lightly doped Si/SiO$_2$ substrate by using the internal gain mechanism. The gain is realized by the ultrahigh mobility of graphene and the long lifetime of carriers in lightly doped Si/SiO$_2$ substrate. The proposed device structure can detect electrostatic potential as low as ~5 V, which is an order of magnitude lower than that of commercialized product. Meanwhile, the graphene based electrostatic sensor possesses fast response speed of ~2 μs and its performance could be efficiently tuned by applying gate or bias. Moreover, the sensor features simplicity in fabricate process and is fully compatible with silicon technology, which would greatly facilitate its industrial application.

**Results**

The schematic of graphene based electrostatic sensor is shown in Fig. 1(a), which contains a two terminal graphene transistor on a lightly n-doped Si/SiO2 substrate. When a negative static electricity source approaches the device, the carriers in Si substrate uncovered by graphene will be redistributed under electrostatic field, while those underneath graphene (which is grounded) would not be affected because of the electrostatic shielding effect. The electrons in Si substrate uncovered by graphene would either move to the bottom of the bulk Si, or drift along the $SiO_2$/Si interface to the region under graphene. The later ones could be trapped by the interface states[20,21] of $SiO_2$/Si, and therefore gate the graphene channel and increase the holes concentration through capacity coupling. Due to the ultrahigh mobility of graphene and long carriers trapping time at the interface, an effective gain[22,23] or amplification will be introduced because holes in the graphene can circulate many times during the trapping time of electron at the interface, namely interfacial amplification[21]. The gain is defined as $G = \tau_l / \tau_t$, here, $\tau_l$ is the trapping time of carriers at the $SiO_2$/Si interface, $\tau_t$ is the transit time in the graphene channel, which is analog to the case of photoconductivity. Such an interfacial amplification process behaves like a built-in amplifier in the system. Remarkably, it would not increase the circuit noise level, so that the sensitivity of the device could be dramatically enhanced.

According to the above mechanism, we prepared a single layer graphene device on 300 nm $SiO_2$/lightly n-doped Si (resistivity of ~50 Ω cm) substrate via electron beam lithography (EBL) and lift off process. The optical image of the device as well as the Raman spectrum is shown in Fig. 1(b), which suggest the monolayer thickness and high quality (no Raman D peak appears[24]) of the graphene sample. The transfer curve of the graphene device is also shown in Fig. 1(b). Evidently, the graphene channel is slightly p-doped and the calculated field-effect mobility is ~6000 $cm^2$/Vs. The electrostatic response of the device is shown in Fig. 1(c), with the electrostatic field switched on and off by inserting a grounded metal plate between the device and the electrostatic source (schematic shown in Supplementary Fig. S1). Here, the standard work distance of 25 mm between the device and static electricity source is adopted. An obvious increase of channel current under negative electrostatic field is observed, indicating the increase of hole concentration in the p-type graphene. For comparison, the electrostatic switching response of the graphene device directly putting on lightly-doped Si substrate, as well as the bare Si channel are also shown in Fig. 1(c). No noticeable electrostatic responses are observed in the above two devices, which suggest that the interfacial amplification does play an important role in our electrostatic sensor. The functionality of the trap states to the gate effect has been verified in the photodetection device employing similar device architecture to our sensor[21]. To further certify the interfacial gain effect of the graphene/$SiO_2$/Si system, we test the photodetection performance of our device (Supplementary Information Figure S2). The results are consistent with previous reports[21] and this indicates that the interfacial amplification is indeed an important factor to the graphene/SiO$_2$/Si based devices. To study the effect of lightly-doped Si on the interfacial amplification process of graphene/SiO$_2$/Si system, electrostatic switching characteristics of the graphene device on different resistivity n-doped Si/SiO$_2$ substrate was studied (see Supplementary Fig. S3). The electrostatic response is enhanced with the increase of the resistivity of n-doped Si (Figure 1(d)), which is consistent with the longer trapping time or lifetime of carriers in lightly-doped Si[21,25]. When the Si substrate is grounded, the output of the devices will be dominated by the response from bare graphene. Nevertheless, the device does not show noticeable response to the electrostatic field after the Si substrate is grounded. This result indicates that the static electricity induced charge transport in bare graphene transistor is negligible.

Electrostatic switching characteristics of the sensor under varying electrostatic potential and work distance were carried out (see Supplementary Fig. S4). Figure 2(a) shows the electrostatic potential dependence of the electrostatic response of our device upon different work distance. The electrostatic potential was calibrated by a commercial electrostatic field meter at standard work distance of 25 mm (see Supplementary Fig. S5). Notably, the electrostatic response increases almost linearly with electrostatic potential, which is consistent with the factor that more electrons drift to the region underneath graphene and trapped by the interface states. The dependence of electrostatic response on the work distance is shown in Fig. 2(b), and it reveals that the response depends linearly on the distance in the dual logarithmic coordinate diagram. The dependence can be described well by the Allometric function y=ax$^b$ (solid lines in Fig. 2(b)). The fitting value of b is ~-0.9, close to -1, which is consistent with the reciprocal relationship between potential (φ) induced by electrostatic source and distance (r): $\varphi = \frac{kQ}{r}$.

Sensitivity is an important parameter for electrostatic sensor, and high sensitivity means not only the detection of weak electrostatic signals, but also the ability to detect electrostatic signals with ultra-long distance. Here, we measured the weak electrostatic signals through increasing work distance and compared the detection capabilities of our device with the commercial product. As shown in Fig. 2(c), the detection limit of electrostatic potential of our device is ~5 V, which is reduced by an order of magnitude as compared to the commercial detector (~50 V). This demonstrates the potential of our device for highly sensitive or ultra-long distance electrostatic signal detection. It should be noted that the detection limit of our electrostatic sensor could be further improved by minimizing the noise level in the graphene channel, which is ~0.5 μA in the current device.

The response speed of the device was characterized by using AC square wave signal with frequency of 100 kHz as the power source of the electrostatic generator. As shown in Fig. 2(d), the device exhibits fast response, with both the rise and fall response time < 2 μs. The fall time ($10^{-6}$ s) could be approximately considered as the lifetime of electrons in lightly n-doped Si, while the transit time $\tau_t$ in the graphene channel is in the order of $10^{-10}$ s at 1 V bias for our device, leading to the gain

($G = \tau_l/\tau_t$ ) of ~$10^4$. Such an ultrahigh gain or amplification is the reason why strong electrostatic response is observed in our device, while the graphene on Si or bare Si device have negligible response (Fig. 1(c)). It is worth noting that, the electrostatic response of the device presents a saltation at the switch moment of static field. This may because the electrostatic field has an abrupt change during the fast switching of power source.

We also investigated the transfer characteristics of the device under different electrostatic potential in a top gate configuration by using 80 nm $Al_2O_3$ as the dielectric layer, as shown in Fig. 3(a). The transfer curve shifts forward under the negative static electricity, indicating a p-doping effect of graphene, which consistent with the previous discussion. The shift of the Dirac point ($\Delta V_{Dirac}$) of the device as a function of electrostatic potential is shown in Fig. 3(b). The nearly linear relation suggests that the number of electrons trapped at the interface or underneath graphene increases almost linearly with the electrostatic potential. Figure 3(c) shows the variation of the electrostatic response with the top gate voltage under different electrostatic potential. The electrostatic response can be reversed in sign and can even be switched off electrically by tuning the gate. Such phenomenon is due to the change of the type of conducting carrier in graphene from hole to electron and its number also changes with applied top gate voltage[21]. Furthermore, the linear dependence of the electrostatic response on bias under different static field is also shown in Fig. 3(d). The above results demonstrate the versatile tunability of our electrostatic sensor, which is an attractive feature in practical applications.

To prove the capability of our graphene based electrostatic sensor, different objects that have different types and amount of static electricity are tested at the same work distance (25mm) and the results are shown in Fig. 4. Glass rod and hand normally carry slight positive charges, while polyethylene terephthalate (PET) naturally carries a large amount of positive charges. As a result, the device exhibits a negative response when the above objects approach it. The magnitude of the electrostatic responses represent different amount of charges carried by the objects. On the other hand, the electrostatic response is positive for a pen and rubber rod, because they carry negative charges after rubbing with hair. The high sensitivity of the sensor to different types and amounts of static electricity demonstrates that our devices are applicative to different objects and environments.

## Conclusion

We have demonstrated a new concept of electrostatic sensor constructed by depositing a graphene transistor onto lightly-doped Si/SiO$_2$ substrate. An interfacial amplification process is realized by the ultrahigh mobility of graphene and long trapping time of carriers in lightly n-doped Si, resulting in an ultrahigh gain of ~10$^4$. The proposed sensor shows fast response of ~2 μs and excellent sensitivity to electrostatic potential as low as ~5 V, which is improved by an order of magnitude as compared to commercial product. More importantly, the sensor does not require any external amplifiers, which would greatly simplify the design of integrated micro sensors.

## Methods

**Device fabrication and characterization.** The mechanically exfoliated single layer graphene was deposited on the top of 300 nm thick silicon oxide/lightly doped silicon ($SiO_2$/Si) substrate, the source and drain electrodes (Ni (5 nm)/Au (50 nm)) were prepared by electron-beam lithography (FEI, FP2031/12 INSPECT F50), thermal evaporation (TPRE-Z20-IV), and lift-off processes. Electrical curve and electrostatic response characteristics of the devices were measured using a Keithley 2612 analyzer. All the measurements were performed under ambient conditions at room temperature. A negative electrostatic generator (SG-81103) was employed to attain static electricity, and the values were calibrated by a commercial electrostatic field meter (FMX-003) at a standard work distance of 25 mm. In the response time measurement, a signal generator (DG 4162) was used to obtain AC electric field signal with high frequency of 100 kHz. A digital storage oscilloscope (Tektronix TDS 1012, 100 MHz/1GS/s) was used to measure the transient response.

**Reference:**

1. Xu, C., Wang, S., Tang, G., Yang, D. & Zhou, B. Sensing characteristics of electrostatic inductive sensor for flow parameters measurement of pneumatically conveyed particles. *J. Electrost.* **65,** 582-592 (2007).
2. Chubb, J. The measurement of atmospheric electric fields using pole mounted electrostatic fieldmeters. *J. Electrost.* **72,** 295-300 (2014).
3. Zhang, W., Yan, Y., Yang, Y. & Wang, J. Measurement of flow characteristics in a bubbling fluidized bed using electrostatic sensor arrays. *IEEE T. Instrum. Meas.* **65,** 703-712 (2016).

4. Huang, J., Wu, X., Wang, X., Yan, X. & Lin, L. A novel high-sensitivity electrostatic biased electric field sensor. *J. Micromech. Microeng.* **25,** 095008 (2015).

5. Ma, J. & Yan, Y. Design and evaluation of electrostatic sensors for the measurement of velocity of pneumatically conveyed solids. *Flow Meas. Instrum.* **11,** 195–204 (2000).

6. Chen, X. et al. Thermally driven micro-electrostatic fieldmeter. *Sensor. Actuat. A* **132,** 677-682 (2006).

7. Seaver, A. E. Moving ground plane electrostatic fieldmeter measurements. *J. Electrost.* **42,** 185-192 (1997).

8. Tajdari, T., Rahmat, M. F. & Wahab, N. A. New technique to measure particle size using electrostatic sensor. *J. Electrost.* **72,** 120-128 (2014).

9. Shao, J., Krabicka, J. & Yan, Y. Velocity measurement of pneumatically conveyed particles using intrusive electrostatic sensors. *IEEE T. Instrum. Meas.* **59,** 1477-1484 (2010).

10. Novoselov, K. S. et al. Electric field effect in atomically thin carbon films. *Science* **306,** 666-669 (2004).

11. Wang, H. et al. Integrated circuits based on bilayer MoS2 transistors. *Nano Lett.* **12,** 4674-4680 (2012).

12. Buscema, M. et al. Fast and broadband photoresponse of few-layer black phosphorus field-effect transistors. *Nano Lett.* **14,** 3347-3352 (2014).

13. Novoselov, K. S. et al. Two-dimensional gas of massless Dirac fermions in graphene. *Nature* **438,** 197-200 (2005).

14. He, Q., Wu, S., Yin, Z. & Zhang, H. Graphene-based electronic sensors. *Chem. Sci.* **3,** 1764-1772 (2012).

15. Yuan, W. & Shi, G. Graphene-based gas sensors. *J. Mater. Chem. A* **1,** 10078-10091 (2013).

16. Robinson, J. T., Perkins, F. K., Snow, E. S., Wei, Z. & Sheehan, P. E. Reduced graphene oxide molecular sensors. *Nano Lett.* **8,** 3137-3140 (2008).

17. Dong, X., Shi, Y., Huang, W., Chen, P. & Li, L. Electrical detection of DNA

hybridization with single-base specificity using transistors based on CVD-grown graphene sheets. *Adv. Mater.* **22,** 1649-1653 (2010).

18. Liu, Y., Dong, X. & Chen, P. Biological and chemical sensors based on graphene materials. *Chem. Soc. Rev.* **41,** 2283-2307 (2012).

19. Ang, P. K., Chen, W., Wee, A. T. S. & Loh, K. P. Solution-gated epitaxial graphene as pH sensor. *J. Am. Chem. Soc.* **130,** 14392-14393 (2008).

20. Schroder, D. K. Surface voltage and surface photovoltage: History, Theory and Applications. *Meas. Sci. Technol.* **12,** R16–R31 (2001).

21. Guo, X. et al. High-performance graphene photodetector using interfacial gating. *Optica* **3,** 1066-1070 (2016).

22. Konstantatos, G. et al. Hybrid graphene–quantum dot phototransistors with ultrahigh gain. *Nat. Nanotech.* **7,** 363-368 (2012).

23. Liu, Y. et al. Planar carbon nanotube–graphene hybrid films for high-performance broadband photodetectors. *Nat. Commun.* **6,** 8589 (2015).

24. Ni, Z. H. et al. On resonant scatterers as a factor limiting carrier mobility in graphene. *Nano Lett.* **10,** 3868-3872 (2010).

25. Cuevas, A. & Macdonald, D. Measuring and interpreting the lifetime of silicon wafers. *Sol. Energy* **76,** 255–262 (2004).

**ACKNOWLEDGEMENTS**

This work was supported by NSFC (61422503 and 61376104), the open research funds of Key Laboratory of MEMS of Ministry of Education (SEU, China), and Research and Innovation Project for College Graduates of Jiangsu Province No. KYLX15_0111.

# Figure captions

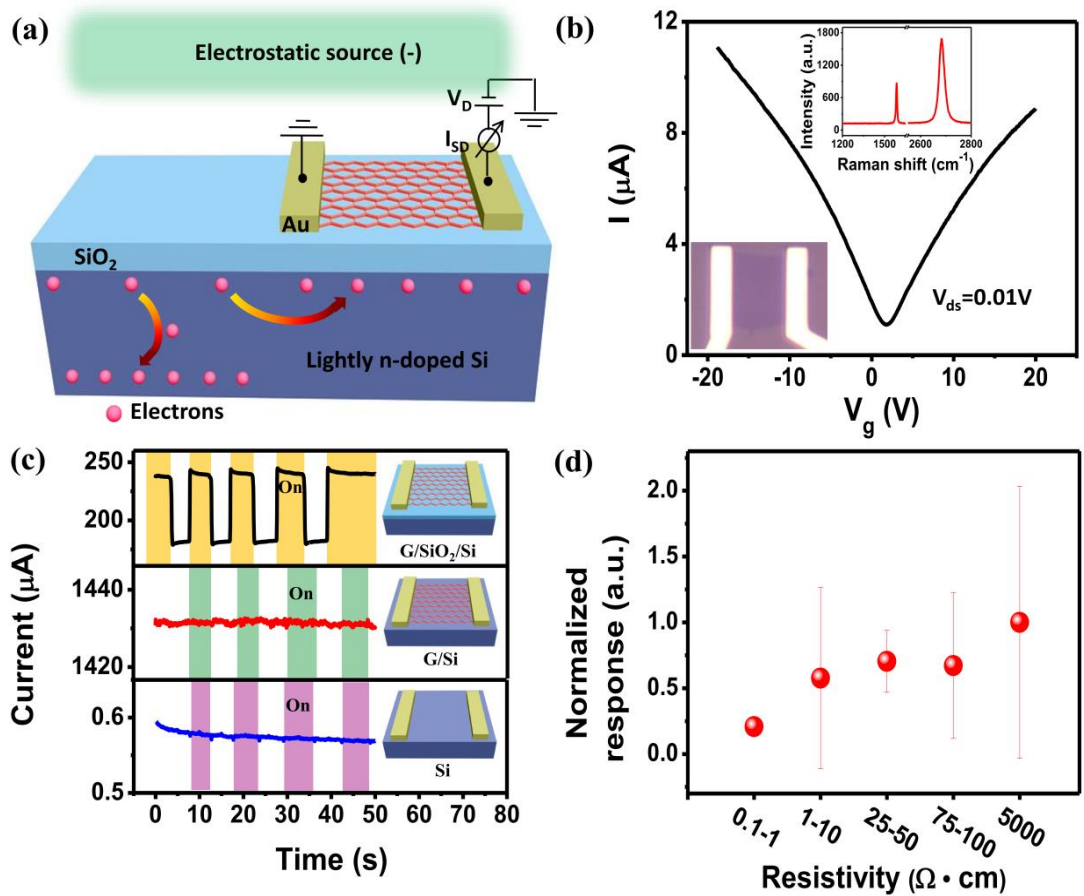

**Figure 1 | Schematic of graphene based electrostatic sensor.** (**a**) Schematic diagram of the graphene-based electrostatic sensor. (**b**) Transfer curve of the device on lightly doped $SiO_2/Si$. The inset is the optical image and Raman spectrum of the graphene channel. (**c**) Electrostatic switching characteristics of the graphene device on lightly doped $SiO_2/Si$, as well as graphene device on lightly doped Si and bare Si device. (**d**) The normalized electrostatic response as a function of the Si doping level of the $SiO_2/Si$ substrate. The response is normalized by the mobility and the width of the graphene device.

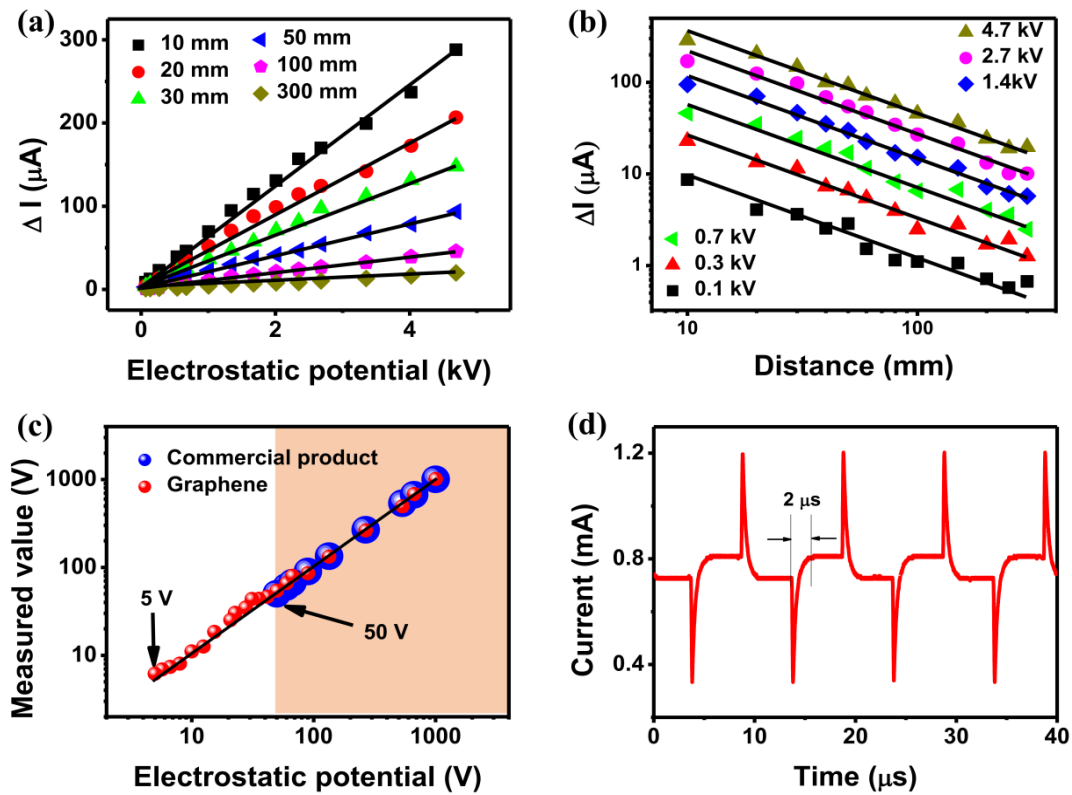

**Figure 2 | Performance of the graphene based electrostatic sensor.** (**a**) The linear electrostatic response of the graphene device to the negative static electricity at different work distance with $V_D = 1$ V. (**b**) The electrostatic response as a function of the distance between the graphene device and negative electrostatic source. The solid line is the fitting curve of the experimental data by using the Allometric function $y=ax^b$. (**c**) Comparison of electrostatic detection sensitivity of the graphene and a commercial product. (**d**) Electrostatic switching characteristics of the graphene device under high frequency of 100 kHz.

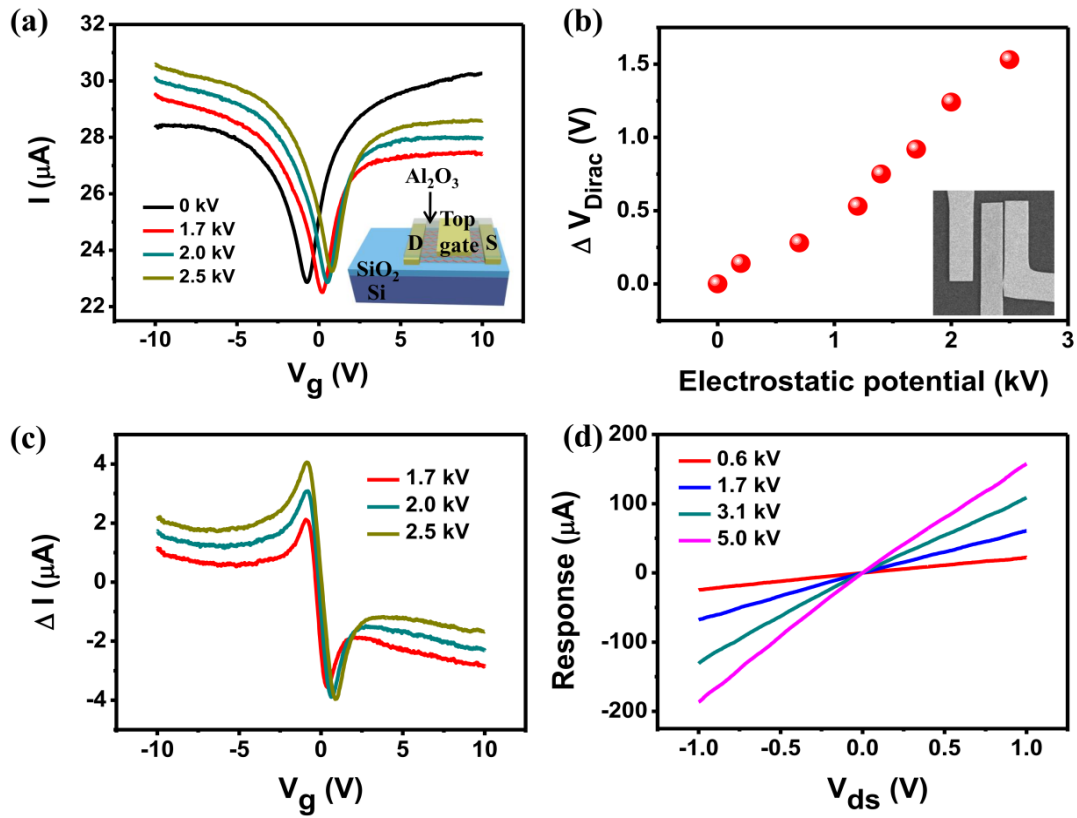

**Figure 3 | Gate- and bias- modulated electrostatic response.** (**a**) I-$V_g$ characteristics of the device under varying static electricity; the inset is schematic setup of top-gated graphene device. (**b**) The shift of the Dira c point as a function of the static electricity; the inset is scanning electron micrograph of top-gated graphene device. (**c**) The extracted gate dependence of electrostatic response from the curves in (a). (**d**) Electrostatic response of the device as a function of bias ($V_{ds}$) under different static electricity.

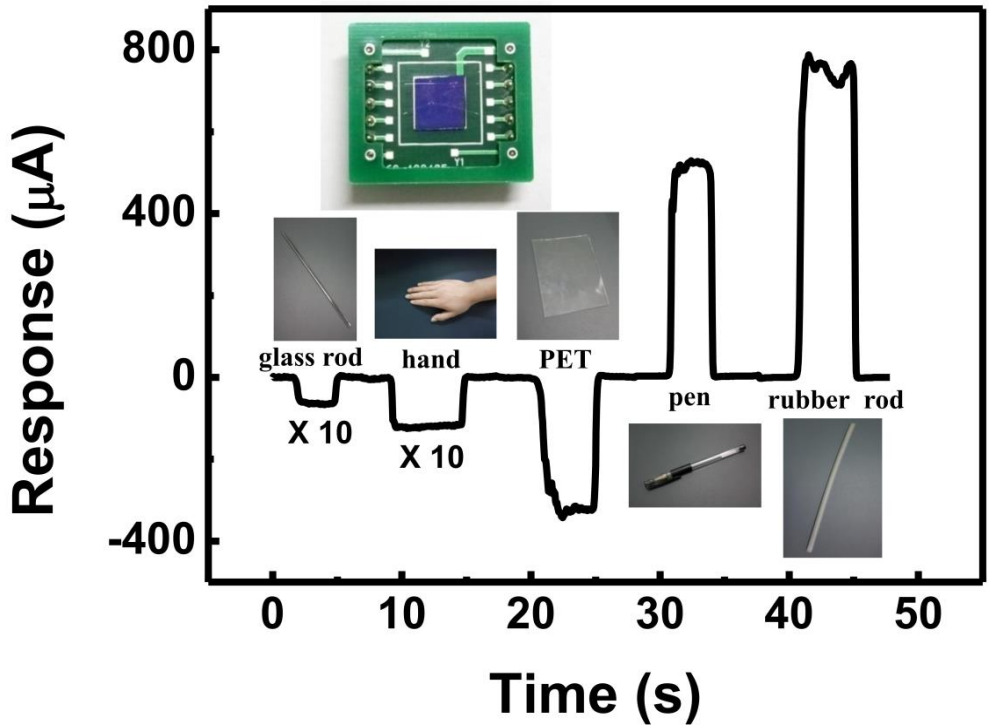

**Figure 4 | Demo of electrostatic sensor.** Electrostatic response of the sensor (photo displayed in the inset) to different objects, showing negative response for positively charged objects (glass rod, hand and PET) and positive response for negatively charged objects (pen and rubber rod).

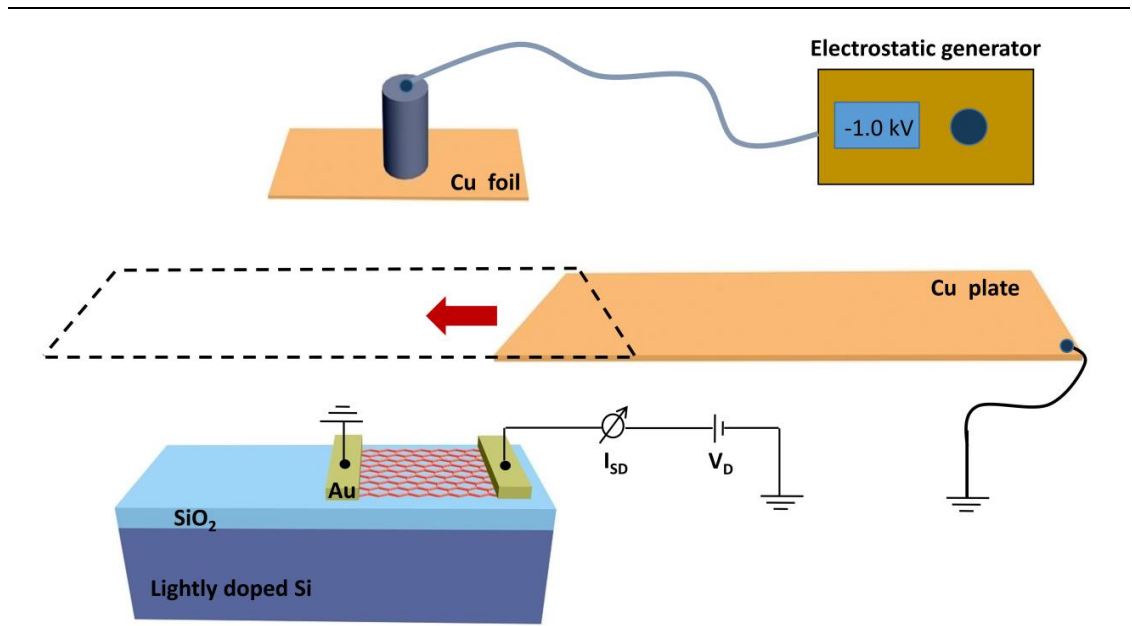

**Supplementary Figure S1 | Schematic of the electrostatic response measurement of the electrostatic field sensor.** The negative charge generated by the electrostatic generator is adsorbed on the copper foil, making it a negative electrostatic source and maintaining a stable potential. The electrostatic field is switched by inserting a grounded copper plate between the device and the electrostatic source.

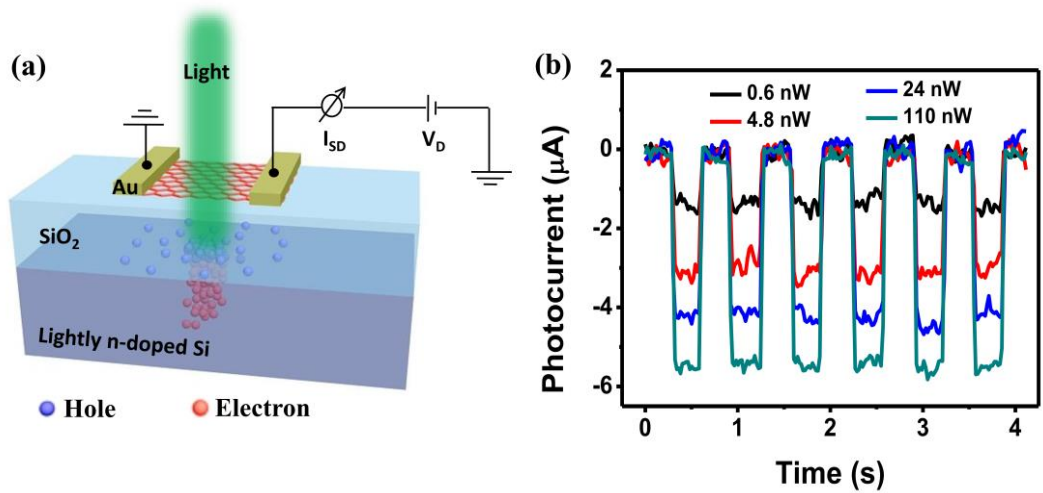

**Supplementary Figure S2 | Schematic of the photodetector based on graphene/SiO$_2$/lightly n-doped Silicon system.** (a) Incident light creates electron–hole pairs in the lightly n-doped silicon. Electrons remain in the silicon, while holes are accumulated and trapped at SiO$_2$/Si interface. (b) Photoswitching characteristics of the graphene photodetector under varying light power. The observable photocurrent is obtained even at low power down to 0.5 nW, which is due to high gain of interfacial amplification.

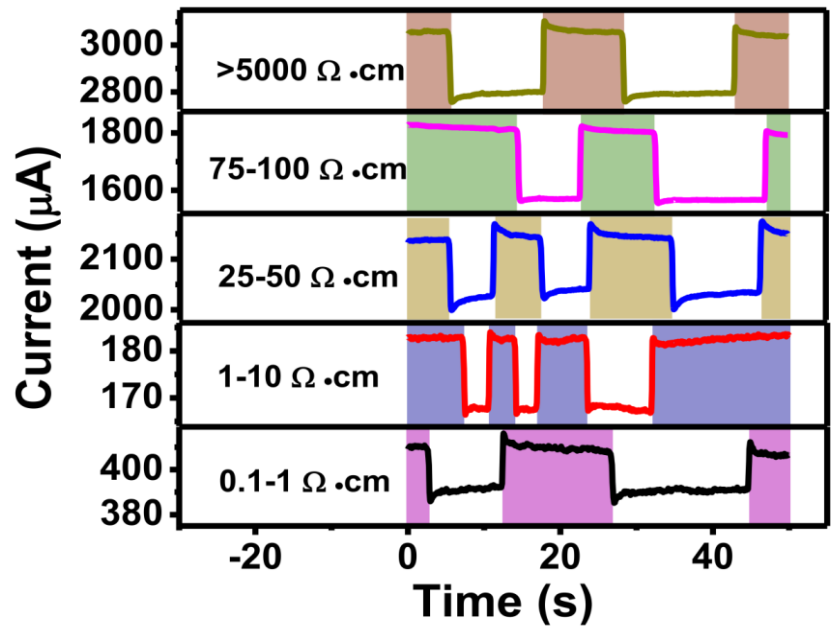

**Supplementary Figure S3 | Electrostatic response of different devices on SiO2/n-doped Si substrate.** Electrostatic switching characteristics of the graphene device on different $SiO_2$/n-doped Si substrate at standard work distance of 25 mm, showing positive response for the negatively charged objects.

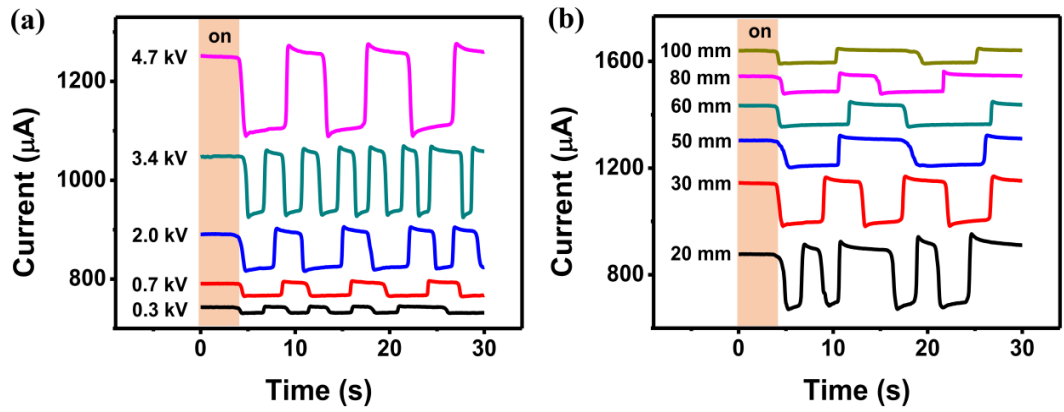

**Supplementary Figure S4 | Electrostatic response characteristics.** Electrostatic switching characteristics of the sensor under varying electrostatic potential (a) and work distance (b). The current in both figures were plotted with vertical offset.

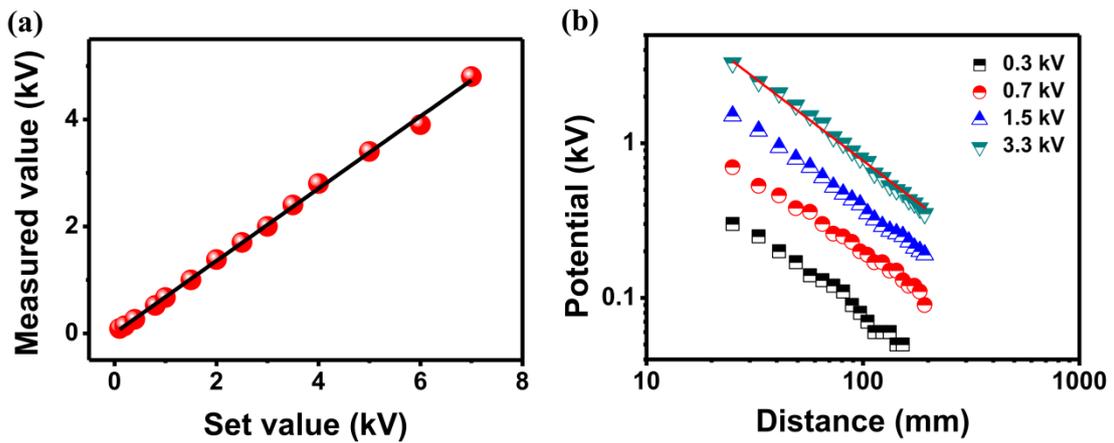

**Supplementary Figure S5 | The calibration of electrostatic potential by using a commercial electrostatic field meter.** (a) The electrostatic potential of copper foil were measured by a commercial electrostatic field meter at standard work distance of 25 mm, and the measured value is linear with the set value of the electrostatic generator. The slope of the fitting line is 0.67, which implies the presence of electrostatic leakage. The electrostatic potential used in the text is the value after calibration. (b) The work distance dependence of measured electrostatic potential was carried out. The solid line is the fitting curve of the experimental data by using the Allometric function $y=ax^b$ and the fitting value of b is ~ -1, which is consistent with the reciprocal relationship between potential and distance.